\documentclass{Interspeech}

\usepackage{amsmath,graphicx,multirow}
\usepackage{bbding}
\usepackage{booktabs}
\usepackage{cite}
\usepackage{verbatim}

\usepackage[utf8]{inputenc}
\usepackage[titletoc,title]{appendix}
\usepackage{makecell}
\usepackage{tipx}

\newcommand{\xmark}{\ding{55}}
\newcommand{\cmark}{\ding{51}}

\usepackage{color}
\usepackage{multirow}
\usepackage{hhline}
\usepackage{subfigure}
\usepackage{pifont}
\usepackage{bm}
\usepackage{amsmath,amsfonts,amssymb,mathtools}

\usepackage{graphicx,float}

\interspeechcameraready

\title{On-the-fly Routing for Zero-shot MoE Speaker Adaptation of Speech Foundation Models for Dysarthric Speech Recognition}

\author[affiliation={1}]{Shujie}{Hu}
\author[affiliation={2*}]{Xurong}{Xie}
\author[affiliation={3}]{Mengzhe}{Geng}
\author[affiliation={1}]{Jiajun}{Deng}
\author[affiliation={1}]{Huimeng}{Wang}
\author[affiliation={1}]{Guinan}{Li}
\author[affiliation={1}]{Chengxi}{Deng}
\author[affiliation={1}]{Tianzi}{Wang}
\author[affiliation={1}]{Mingyu}{Cui}
\author[affiliation={1}]{Helen}{Meng}
\author[affiliation={1*}]{Xunying}{Liu}

\affiliation{}{The Chinese University of Hong Kong}{Hong Kong SAR, China}
\affiliation{Institute of Software}{Chinese Academy of Sciences}{China}
\affiliation{}{National Research Council Canada}{Canada}
\email{\{sjhu,xyliu\}@se.cuhk.edu.hk, xurong@iscas.ac.cn}

\usepackage{comment}

\begin{document}

\maketitle
\renewcommand{\thefootnote}{*}%
\footnotetext{Corresponding author.}%
\renewcommand{\thefootnote}{\arabic{footnote}}%
\begin{abstract}
This paper proposes a novel MoE-based speaker adaptation framework for foundation models based dysarthric speech recognition. This approach enables zero-shot adaptation and real-time processing while incorporating domain knowledge. Speech impairment severity and gender conditioned adapter experts are dynamically combined using on-the-fly predicted speaker-dependent routing parameters. KL-divergence is used to further enforce diversity among experts and their generalization to unseen speakers. Experimental results on the UASpeech corpus suggest that on-the-fly MoE-based adaptation produces statistically significant WER reductions of up to 1.34\% absolute (6.36\% relative) over the unadapted baseline HuBERT/WavLM models. Consistent WER reductions of up to 2.55\% absolute (11.44\% relative) and RTF speedups of up to 7 times are obtained over batch-mode adaptation across varying speaker-level data quantities. The lowest published WER of 16.35\% (46.77\% on very low intelligibility) is obtained.
\end{abstract}

\section{Introduction}

Despite the rapid progress of ASR technologies targeting normal and healthy users, their application to those suffering from speech disorders, such as dysarthria, remains a challenging task to date\cite{sehgal2015model,xiong2018deep,liu2021recent,geng2022speaker, hu2022exploring, yue2022acoustic, hu23b_interspeech, hu2022exploiting, 10584335, hu2024structured}.
Dysarthric speech brings challenges on all fronts to current deep learning based ASR technologies predominantly targeting healthy users: \textbf{1) substantial mismatch} against typical voices; \textbf{2) data scarcity} \cite{liu2021recent, 10584335}; and \textbf{3) large speaker-level diversity} \cite{smith1987temporal} among dysarthric talkers, including accent or gender, and speech pathology severity. 
Such heterogeneity among dysarthric speakers not only complicates the training or fine-tuning of speaker-independent (SI) ASR systems on this data but also hinders their effective personalization to individual users’ voices. 
These challenges are further compounded in the fine-tuning of self-supervised learning (SSL) speech foundation models (SFMs) \cite{baevski2020wav2vec, chen2022wavlm, hsu2021hubert} with their massive parameter counts.
Despite the growing prominence of SFMs in ASR research, limited studies have investigated speaker adaptation of SFMs for dysarthric speech recognition \cite{baskar2022speaker, hu2024structured, jiang24b_interspeech}. 
\par
Recent studies on SFM adaptation \cite{thomas2022efficient, 10023274, 10096330, 10193427} have primarily focused on normal speech, with most approaches utilizing a single adapter for transfer learning. 
In contrast, Mixture of Experts (MoE) methods \cite{jacobs1991adaptive, jordan1994hierarchical, shazeer2017outrageously} demonstrate superior effectiveness addressing data heterogeneity such as speaker-level diversity. 
Individual experts develop specialized capabilities to handle specific data distributions, while their diversity enables comprehensive coverage and generalization to unseen data.
These MoE methods have been extensively applied in large language models (LLMs) \cite{jiang2024mixtral, dai2024deepseekmoe, sun2024hunyuan} and widely adopted in speech recognition, where they have been integrated into end-to-end Transformer or Conformer \cite{you21_interspeech, bai22_interspeech, 10389798, 10096227, perez20_interspeech} as well as pre-trained SFMs \cite{zhao24d_interspeech}. 
\par
However, existing research on MoE approaches has predominantly focused on typical speech. When applying MoE to dysarthric speaker adaptation, three significant issues emerge: \textbf{a)} 
Mobility issues of dysarthric speakers hinder large-scale speech data collection, resulting in data sparsity and speaker bias. 
Such issues severely restrict the generalizability to unseen speakers;
\textbf{b)} Batch-mode unsupervised test-time adaptation introduces substantial processing delays due to its two-stage process: pseudo-label generation followed by speaker-dependent (SD) parameter updates.
The resulting high latency imposes a significant physical burden on dysarthric users during interactions; 
and \textbf{c)} The pathological nature of dysarthric speech necessitates the incorporation of domain knowledge to ensure both diverse expert specialization and comprehensive coverage of the MoE.
\par
To address these challenges, we propose a novel MoE-based speaker adaptation framework for SFM based dysarthric speech recognition.
This approach enables \textbf{zero-shot} adaptation and \textbf{real-time} processing while effectively incorporating \textbf{domain knowledge}.
Specifically, feature-driven routing networks are designed to generate homogeneous SD routing parameters on the fly, enabling \textbf{a) zero-shot} adaptation and \textbf{b) real-time} processing. In addition, \textbf{c) domain knowledge}, such as severity and gender information, is incorporated by initializing each expert with severity and gender conditioned adapter parameters from adaptive training \cite{hu2024structured}, allowing experts to focus on distinct severity and gender groups.
These severity and gender labels are further utilized in classification tasks to better capture dysarthric speaker characteristics. 
Additionally, a Kullback-Leibler (KL) divergence loss is introduced during training to further enforce diversity among experts and their generalization to unseen speakers's data.
\par
The main contributions of our work are summarized below: \\
\textbf{1) {Novelty:}} To the best of our knowledge, this paper is the first to investigate on-the-fly MoE-based speaker adaptation for dysarthric speech recognition, whereas prior efforts have primarily focused on typical speech \cite{zhao24d_interspeech}.
Our method addresses the three major challenges outlined earlier: \textbf{a)} while previous methods lack the ability to adapt to unseen speakers, our \textbf{zero-shot} approach achieves greater generalization and extends applicability; \textbf{b)} our \textbf{on-the-fly} router achieves real-time speaker adaptation, which is more efficient compared to previous two-stage batch-mode methods
\cite{zhao24d_interspeech, hu2024structured}; and \textbf{c)} compared to previous approaches focusing on speaker identity only \cite{zhao24d_interspeech}, we leverage \textbf{domain knowledge} to better model dysarthric speakers. \\
\textbf{2) {Performance:}} Experimental results on the UASpeech \cite{kim2008dysarthric} dysarthric corpus suggest that \textbf{i)} the proposed on-the-fly MoE-based adaptation approach produces statistically significant word error rate (WER) reductions of up to \textbf{1.34\%} absolute (\textbf{6.36\%} relative) over the baseline SI HuBERT and WavLM models. \textbf{ii)} Consistent WER reductions of up to \textbf{2.55\%} absolute (\textbf{11.44\%} relative) and \textbf{iii)} real-time factor (RTF) speedups of up to \textbf{7 times} are obtained over batch-mode adaptation across varying speaker-level data quantities.
Further, \textbf{iv)} the lowest published WER of \textbf{16.35\%} (\textbf{46.77\%} on very low intelligibility) is obtained after cross-system multi-pass rescoring \cite{10584335}. \\
\textbf{3) {Analysis:}} Heatmap visualization intuitively reveals that the on-the-fly predicted SD routing parameters exhibit more \textbf{consistent} and \textbf{interpretable} speech impairment severity centric features than those obtained without domain knowledge.

\begin{figure}
    \centering
    \includegraphics[width=0.95\linewidth]{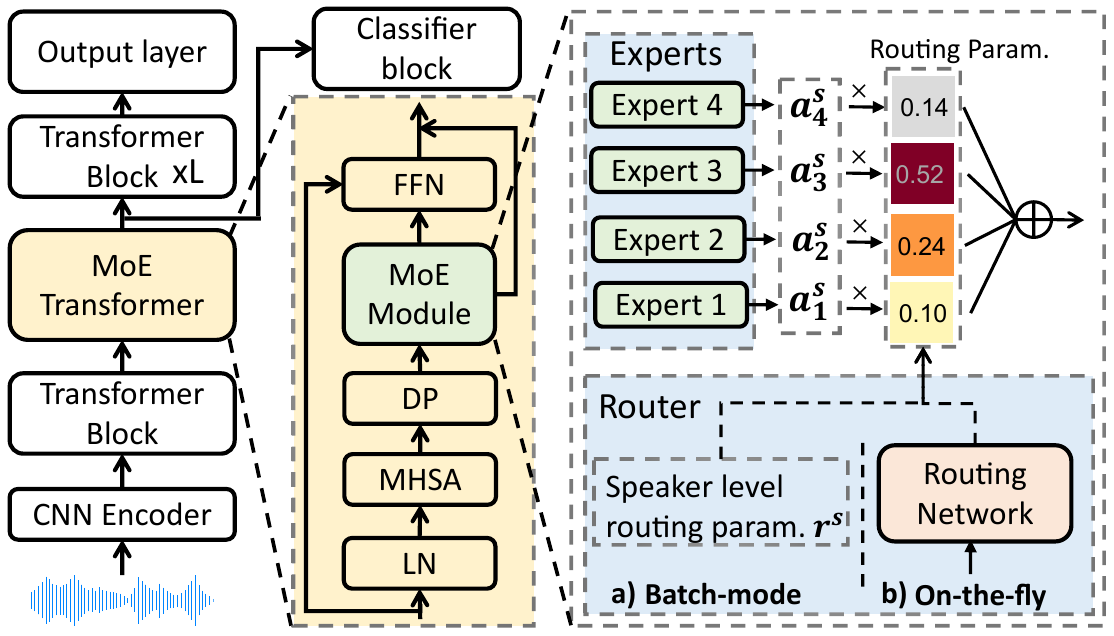}
    \vspace{-0.1cm}
    \caption{Architecture of MoE-based adaptation on SFM, where the routing parameters are derived from either \textbf{a)} speaker-dependent parameters in batch mode; or \textbf{b)} an on-the-fly routing network. ``LN'', ``MHSA'', ``DP'' and ``FFN'' are layernorm, multi-head self-attention, dropout and feedforward.}
    \label{fig:sat_model}
    \vspace{-0.6cm}
\end{figure}


 \begin{figure*}[h]
    \centering
    \includegraphics[width=0.92\linewidth]{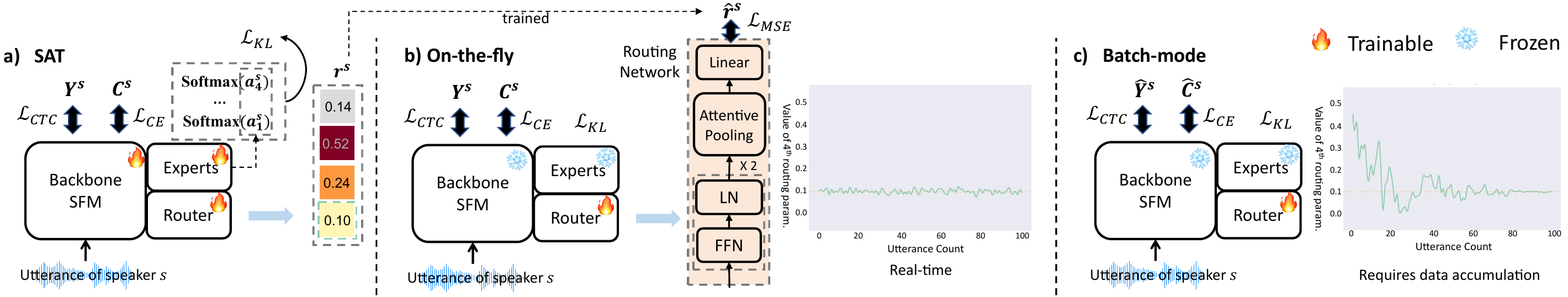}
    \vspace{-0.2cm}
    \caption{Examples of \textbf{on-the-fly (a) \& b)}) and \textbf{batch-mode (a) \& c)}) MoE-based speaker adaptation on the SFM.
    Routing parameters $\bm{r}^{s}$ in a) serve as SD parameters, while experts are shared by all speakers.
    The line charts in b) and c) illustrate the variation in a specific expert's routing parameters as a function of utterance count.}
    \label{fig:sat_tta}
    \vspace{-0.6cm}
\end{figure*}

\vspace{-3pt}
\section{Batch-Mode MoE Speaker Adaptation}\label{batch_mode_sec}
\par
\noindent
\textbf{Backbone Speech Foundation Models (SFMs):} SSL speech foundation models such as Wav2vec2.0 \cite{baevski2020wav2vec}, HuBERT \cite{hsu2021hubert}, and WavLM \cite{chen2022wavlm} share similar Transformer-based backbones. For example, HuBERT contains three main components: 1) a multi-layer CNN-based feature encoder; 2) an $L$-layer transformer-based context network with a projection layer; and 3) a $k$-means quantization module. In this paper, we fine-tune the pre-trained HuBERT and WavLM with a CTC decoder.
\par
\noindent
\textbf{MoE Architecture:} As shown in Fig. \ref{fig:sat_model}, the MoE module is integrated into the 2$^{nd}$ Transformer block, positioned between the feedforward layer and the dropout module. Residual Adapter Blocks (RAB) \cite{hu2024structured} act as expert network modules. All speakers share experts, while the router uses SD learnable parameters in batch mode. For speaker $s$, let $N$ and $\bm r^{s}\in\mathbb{R}^{N}$ denote the number of experts and the SD routing parameters. The adapted hidden outputs of the MoE module are given as $\bm{h}^{s}=\sum_{i=1}^{N}{r}_{i}^{s}\bm{a}_{i}^{s}$, where $\bm{a}_{i}^{s}$ denotes the outputs of the $i$-th expert for speaker $s$.
\par
\noindent
\textbf{Multi-task Learning:} To enforce diversity among experts and their generalization to unseen speakers's data,
\textbf{1)} a Kullback-Leibler (KL) divergence loss $\mathcal{L}_{KL}$ is introduced to penalize similarity between the outputs of different experts:
\vspace{-5pt}
\begin{equation}
    \mathcal{L}_{KL} = -\textstyle \sum_{(i,j) \in [N]^2: i \neq j} D_{KL}(\zeta(\boldsymbol{a}_i^s)||\zeta(\boldsymbol{a}_j^s))
\end{equation}
where $D_{KL}$ is the KL divergence and $[N]^2$ denotes the Cartesian product of the set $\{1, ..., N\}$ with itself.
$\zeta(\cdot)$ is a \text{Softmax} function converting $\bm{a}_i^s$ to a probability distribution\footnote{Alternative settings, such as $\zeta(\bm{a}_i^s)=\mathcal{N}(\bm{a}_i^s, 1)$ were found to degrade performance.}. Furthermore, domain knowledge is incorporated by initializing the experts using adapter parameters from adaptive training \cite{hu2024structured}, providing a robust foundation for specialized expert development.
To better capture dysarthric speaker characteristics, \textbf{2)} an auxiliary classification task\footnote{Three levels of domain knowledge are used: severity, severity-gender, and speaker. Ablation studies are conducted in Sec. 4.3.} with cross-entropy (CE) loss is used. As shown in Fig. \ref{fig:sat_tta}(a), the combined batch-mode MoE-based adaptation cost function is
$
    \mathcal{L}_{B} = \mathcal{L}_{CTC} + \alpha \mathcal{L}_{KL} + \beta \mathcal{L}_{CE}
$,
where $\alpha$ and $\beta$ are empirically set to 5 and 0.1 in this paper.
\par
\noindent
\textbf{Test-Time Adaptation:} Unsupervised test-time adaptation is performed on speaker data without speech transcription or classification labels. The classification label\footnote{Classification task is not performed for speaker-level domain knowledge in test-time adaptation.} $\bm{\hat{C}}^{s}$ for test speaker $s$ is automatically predicted using the spectro-temporal feature based neural network classifiers in \cite{geng23b_interspeech, geng2022speaker}.
The hypothesis supervision $\bm{\hat{Y}}^{s}$ for adaptation is generated by decoding the test data using the baseline unadapted SFMs.
During unsupervised test-time adaptation, as shown in Fig. \ref{fig:sat_tta}(c), the speaker-level routing parameters $\bm{r}^{s}$ are re-initialized and optimized by:
\vspace{-3pt}
\begin{equation}
    \{\hat{\bm{r}}^{s}\} = \underset{\{\bm{r}^{s}\}}{ \arg\min} \{\mathcal{L}_{B}(\bm{\hat{Y}}^{s}, \bm{\hat{C}}^{s} | \bm{X}^{s}; \bm{r}^{s})\} 
\end{equation} \\[-6pt]
where $\bm{X}^{s}$ is the input waveform for speaker $s$. All other parameters of the MoE-based SFM remain frozen.
\par
\noindent
\textbf{Speaker Adaptive Training (SAT):} During supervised training, SAT generates speaker-invariant canonical models that provide a more neutral and robust starting point for unsupervised test-time adaptation compared to standard non-SAT models.
As shown in Fig. \ref{fig:sat_tta}(a), the speaker-level routing parameters are jointly optimized with both the backbone SFM's parameters $\bm \Theta$ and those of the experts $\bm{\Theta}_{e}$ during SAT, given as follows:
\vspace{-3pt}
\begin{equation}
    \scalebox{0.83}{
    $\{\hat{\bm{\Theta}}, \hat{\bm{\Theta}}_{e},\hat{\bm{\theta}}_{S}\} = \underset{\{\bm{\Theta}, \bm{\Theta}_{e}, \bm{\theta}_{S}\}}{\arg\min} \sum_{s \in S} \{\mathcal{L}_{B}(\bm{Y}^{s}, \bm{C}^{s}|\bm{X}^{s}; \bm{\Theta}, \bm{\Theta}_{e}, \bm{\theta}_{S})\}$
    }
\end{equation}\\[-6pt]
where $\bm{\theta}_{S}=\{\bm{r}^{s}\}_{s \in S}$ is the SD parameter sets associated with training data. $\bm{Y}^{s}$ and $\bm{C}^{s}$ denote the ground truth transcription and classification label, respectively.
\par
\noindent
\textbf{Adaptation Data:} As shown in the line chart in Fig. \ref{fig:sat_tta}(c), unsupervised test-time adaptation in batch mode is highly dependent on the utterance count. The routing parameters initially fluctuate significantly and require substantial data accumulation to converge, introducing notable processing delays.

\vspace{-0.2cm}
\section{On-the-fly MoE Speaker Adaptation}
\par
\noindent
\textbf{Routing Network Architecture:} To achieve zero-shot speaker adaptation and reduce latency, a feature-driven routing network is designed to generate homogeneous SD routing parameters (shown in the line chart of Fig. \ref{fig:sat_tta}(b)) on the fly, enabling efficient and real-time speaker adaptation.
As depicted in Fig. \ref{fig:sat_tta}(b), the routing network comprises two alternating feedforward layers and layernorm modules, an attentive pooling block for capturing intra-utterance speaker context, and a final linear layer.
\par
\noindent
\textbf{Attentive Pooling:} To capture both the internal contexts of utterances and their temporal dynamics, attentive statistics pooling \cite{okabe2018attentive} is integrated into the routing network. Let $\hat{\bm{h}}^{s,k}_{t}$ and $T^{s,k}$ respectively denote the normalized hidden outputs at time step $t$ and the frame count for the $k$-th utterance of speaker $s$, respectively. The weighted mean and standard deviation are given as $\bm{\mu}^{s,k} = \sum_{t=1}^{T^{s,k}}\alpha^{s,k}_{t}\hat{\bm{h}}^{s,k}_{t}$ and $
    \bm{\sigma}^{s,k} = \sqrt{\sum_{t=1}^{T^{s,k}}\alpha^{s,k}_{t}\hat{\bm{h}}^{s,k}_{t}\odot \hat{\bm{h}}^{s,k}_{t} - \bm{\mu}^{s,k}\odot\bm{\mu}^{s,k}}
$,
where $\alpha^{s,k}_{t}$ is the normalized attention score at time step $t$, given as:
\vspace{-4pt}
\begin{equation}
    \alpha^{s,k}_{t} = \frac{\exp(\bm{v}^T f(\bm{W}\hat{\bm{h}}^{s,k}_{t} + \bm{b}) + c)}{\sum_{t=1}^{T^{s,k}}\exp(\bm{v}^T f(\bm{W}\hat{\bm{h}}^{s,k}_{t} + \bm{b}) + c)}
\end{equation}\\[-5pt]
where $\bm{W}, \bm{b}, \bm{v}$ and $c$ are the trainable parameters. $f(\cdot)$ denotes the $\text{Tanh}(\cdot)$ function. These statistics are then concatenated as $\bm{z}^{s,k}=[\bm{\mu}^{s,k}; \bm{\sigma}^{s,k}]$ and fed into the subsequent linear layer.
\par
\noindent
\textbf{Multi-task Learning:} The primary objective of the routing network is to minimize the mean squared error (MSE) between the predicted SD routing parameters and their corresponding training targets. The training targets $\hat{\bm{r}}^{s}$ can be obtained 
on the training data of speaker $s$ in a supervised manner within the SAT framework (detailed in Sec. \ref{batch_mode_sec}, and shown in Fig. \ref{fig:sat_tta}(a)). 
Therefore, the overall on-the-fly MoE-based speaker adaptation learning cost function is given as $\mathcal{L}_{O} = \mathcal{L}_{CTC} + \alpha \mathcal{L}_{KL} + \beta \mathcal{L}_{CE} + \gamma \mathcal{L}_{MSE}$, where $\gamma$ is empirically set as 0.5 in this paper.
\par
As shown in Fig. \ref{fig:sat_tta}(b), during on-the-fly speaker adaptation, the routing network $\bm{\Theta}_{P}$ is optimized as: 
\vspace{-4pt}
\begin{equation}
    \{\hat{\bm{\Theta}}_P\} = \underset{\{\bm{\Theta}_{P}\}}{\arg\min} \sum_{s \in S} \{\mathcal{L}_{O}(\bm{Y}^{s}, \bm{C}^{s}, \hat{\bm{\theta}}_{S}|\bm{X}^{s}; \bm{\Theta}_{P}) \}
\end{equation}
the parameters of the backbone SFM and experts are initialized using those obtained after SAT and frozen. During decoding, the SD routing parameters of each test utterance are predicted online and directly applied to the outputs of the experts for test-time on-the-fly adaptation.

\vspace{-0.2cm}
\section{Experiments}
\subsection{Task Description and Experimental Setup}
\vspace{-0.1cm}
UASpeech \cite{kim2008dysarthric} is the largest publicly available dysarthric speech dataset containing 16 dysarthric and 13 control speakers.
It includes 155 common and 300 uncommon words and is further divided into three subset blocks per speaker.
The same 155 common words are used across all blocks, while the uncommon words vary. 
Data from Blocks 1 and 3 of all 29 speakers and Block 2 of 13 control speakers form the training set, while Block 2 of 16 dysarthric speakers is the test set. After silence stripping and speed perturbation based data augmentation \cite{geng2020investigation}, 
the training set comprises 173 hours of audio, with 9 hours for evaluation.
The pre-trained models on UASpeech are 
HuBERT\footnote{huggingface.co/facebook/hubert-large-ls960-ft} and WavLM\footnote{huggingface.co/microsoft/wavlm-large}. 
The settings of RAB-based experts follow \cite{hu2024structured}. 

\begin{table}[h]
    \setlength\tabcolsep{1.9pt}
    \centering
    \caption{Performance comparison between baseline, i-vector, x-vector, RAB-based speaker adaptation and the proposed MoE-based speaker adaptation on HuBERT and WavLM. $\dag$ and $\ast$ denote statistically significant (MAPSSWE \cite{gillick1989some}, $\alpha$ = 0.05) improvements obtained against the SI baseline ASR systems (Sys. 1, 8). ``+'' represents score interpolation, while ``X$\rightarrow$Y'' denotes two-pass rescoring of the N-best outputs from system X by system Y. ``VL/L/M/H'' denotes the speech intelligibility groups ``very low'', ``low'', ``mid'' and ``high''.}
    \vspace{-0.2cm}
    \resizebox{\linewidth}{!}{
    \begin{tabular}{c|c|c|c|c|c|c|c} 
\hline\hline
\multirow{2}{*}{Sys.} & \multirow{2}{*}{Model}  & \multirow{2}{*}{\begin{tabular}[c]{@{}c@{}}Adapt.\\Method\end{tabular}} & \multirow{2}{*}{\begin{tabular}[c]{@{}c@{}}\# SD\\Param.\end{tabular}} & \multirow{2}{*}{\begin{tabular}[c]{@{}c@{}}On The\\Fly\end{tabular}} & \multicolumn{2}{c|}{WER (\%)}                                     & \multirow{2}{*}{RTF}    \\ 
\cline{6-7}
                      &                         &                                                                         &                                                                        &                                                                      & VL / L / M / H                                     & All          &                         \\ 
\hline\hline
1    & \multirow{7}{*}{HuBERT} & SI                                          & -                                                     & \xmark                               & 56.83~ /  22.43~ / 11.86~ / 2.70~                     & 21.03~        & 0.31  \\ 
\cline{1-1}\cline{3-8}
2    &                         & i-vector                                             & -                                                     & \cmark                               & 55.79$^\dag$ / 21.87$^\dag$ / 11.98~ / 2.68~        & 20.69$^\dag$ & 0.36  \\
3    &                         & x-vector                                             & -                                                     & \cmark                               & 55.20$^\dag$ / 21.19$^\dag$ / 11.53~ / 2.58~        & 20.26$^\dag$ & 0.35  \\ 
\cline{1-1}\cline{3-8}
4    &                         & \multirow{2}{*}{RAB}                                 & 4M                                                    & \xmark                               & 54.90$^\dag$ / 19.64$^\dag$ / 9.33$^\dag$~ / 2.46~  & 19.34$^\dag$ & 2.59  \\
5    &                         &                                                      & -                                                     & \cmark                               & 57.97~ / 22.70~ / 12.63~ / 2.71~                      & 21.50~        & 0.32  \\ 
\cline{3-8}
6    &                         & \multirow{2}{*}{MoE}                                 & 160                                                   & \xmark                               & 54.76$^\dag$ / 20.93$^\dag$ / 9.88$^\dag$~ / 2.48~  & 19.75$^\dag$ &  2.39     \\
7    &                         &                                                      & -                                                     & \cmark                               & \textbf{54.42$^\dag$} / \textbf{21.02$^\dag$} / \textbf{9.88$^\dag$~} / 2.59~  & \textbf{19.74$^\dag$} & \textbf{0.39}      \\ 
\hline
8    & \multirow{4}{*}{WavLM}  & SI                                         & -                                                     & \xmark                               & 56.49~ /  22.76~ / 11.57~ / 2.89~                     & 21.06~        & 0.42  \\ 
\cline{1-1}\cline{3-8}
9    &                         & RAB                                                  & 4M                                                    & \xmark                               & 53.60$^\ast$ / 20.13$^\ast$ / 9.90$^\ast$~ / 2.35~  & 19.26$^\ast$ & 3.25  \\ 
\cline{1-1}\cline{3-8}
10   &                         & \multirow{2}{*}{MoE}                                 & 160                                                   & \xmark                               & 53.94$^\ast$ / 20.65$^\ast$ / 10.04$^\ast$ / 2.82~ & 19.65$^\ast$ & 3.17  \\
11   &                         &                                                      & -                                                     & \cmark                               & \textbf{53.92$^\ast$} / \textbf{21.00$^\ast$} / \textbf{10.02$^\ast$} / 2.77~ & \textbf{19.72$^\ast$} & \textbf{0.45}  \\ 
\hline\hline
12   & TDNN                    & LHUC-SAT                                             & 25K                                                   & \xmark                               & 61.62~ / 24.56~ / 15.82~ / 6.50~                      & 24.64~        & -     \\ 
\hline
13   & \multicolumn{4}{c|}{(12$\rightarrow$4) + (12$\rightarrow$7) + (12$\rightarrow$9) + (12$\rightarrow$11)}                                                                                    & \textbf{46.77}~ / 16.62~ / 6.53~~ / 2.64~                       & \textbf{16.35}~        & -     \\
\hline\hline
\end{tabular}
}
    \label{tab:main_2}
    \vspace{-0.1cm}
\end{table}

\begin{figure}[h]
    \centering
    \includegraphics[width=0.48\textwidth]{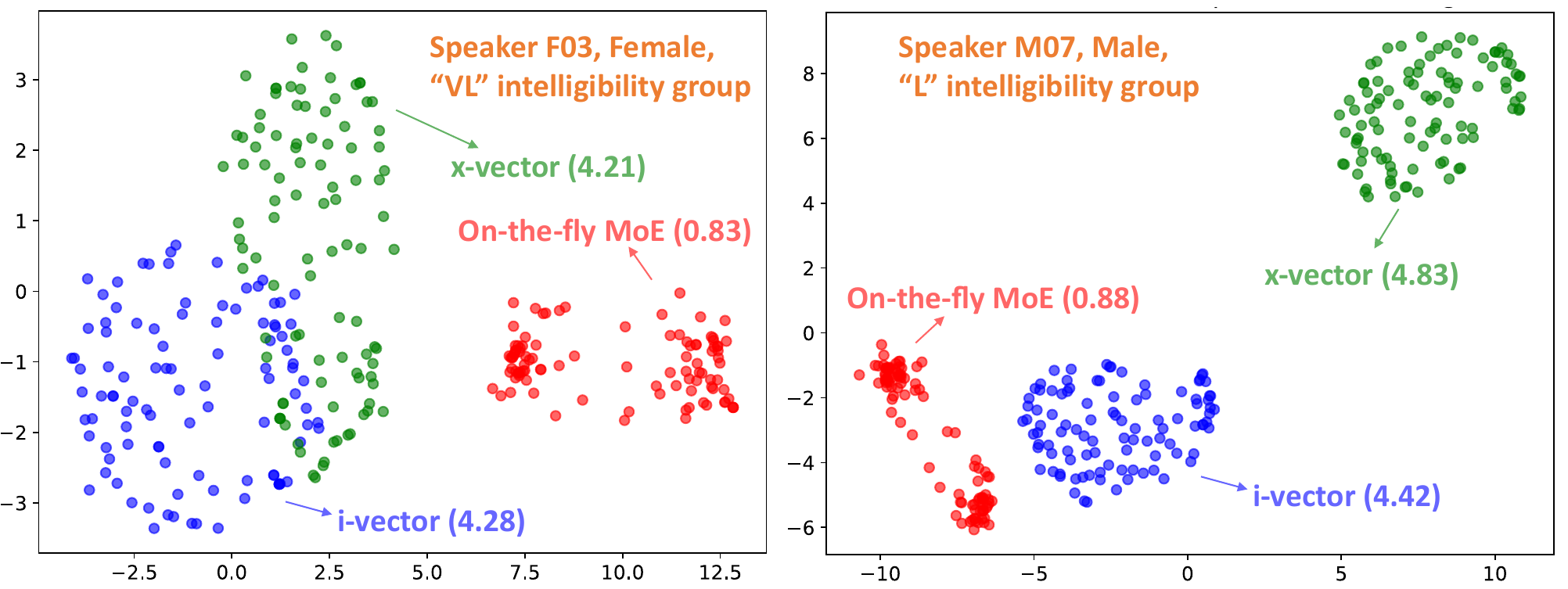}
    \vspace{-0.4cm}
    \caption{T-SNE visualization of the on-the-fly MoE-based, i-vector, and x-vector adaptation. The determinants of their covariance matrices are shown in each bracket.}
    \label{fig:tsne}
    \vspace{-0.7cm}
\end{figure}

\vspace{-0.2cm}
\subsection{Main Results}
\vspace{-0.1cm}
Several trends can be observed from Table \ref{tab:main_2}:
\\ 
\textbf{1) Comparison with the SI baseline, i-vector and x-vector adaptation:} The proposed on-the-fly MoE-based speaker adaptation consistently outperforms these systems with statistically significant WER reductions of up to \textbf{1.34\% absolute (6.36\% relative)} on HuBERT and WavLM (Sys. 7 vs. 1,2,3 \& Sys. 11 vs. 8).
These WER reductions align with the more invariant speaker features produced by on-the-fly MoE in T-SNE visualization compared to i-vectors and x-vectors, as shown in Fig. \ref{fig:tsne}. \\
\textbf{2) Comparison with RAB-based\footnote{The RAB-based method can be considered as a special case of MoE, 
with a single expert and SD parameters as the expert parameters.} methods:} Both batch-mode and on-the-fly MoE-based adaptation achieve comparable average WERs compared to batch-mode RAB-based method \cite{hu2024structured}, while producing lower WERs on the ``VL'' group (Sys. 6 \& 7 vs. 4). Notably, the MoE-based methods require only \textbf{1/25000} of the SD parameters (Sys. 6 vs. 4) and operate at \textbf{1/7} of the RTF (Sys. 7 vs. 4). 
The parameter-heavy RAB method, when applied on-the-fly, underperforms against the SI baseline (Sys. 5 vs. 1). In contrast, the proposed on-the-fly MoE-based adaptation largely outperforms the on-the-fly RAB method with WER reductions of \textbf{1.76\%} absolute (\textbf{8.19\%} relative, Sys. 7 vs. 5). \\
\textbf{3) Comparison with batch-model MoE-based adaptation:} The on-the-fly MoE-based speaker adaptation achieves comparable performance to offline batch-mode MoE-based method, while operating approximately \textbf{7 times} faster in terms of RTF (Sys. 7 vs. 6 \& Sys. 11 vs. 10). \\
\textbf{4) Best performing system:} By combining multiple adapted systems, including LHUC-SAT TDNN, RAB-based adapted SFMs, and the proposed MoE-based adapted SFMs via cross-system rescoring, the lowest published overall WER of \textbf{16.35\%} (\textbf{46.77\%} on very low intelligibility, Sys. 13) is obtained. Finally, the performance of our best system is contrasted against recently published state-of-the-art results in Table \ref{table_sota}.

\vspace{-0.2cm}
\begin{table}[h]
\centering
\caption{WERs of published and our best system on \textbf{UASpeech}
}
\vspace{-0.2cm}
\setlength\tabcolsep{1.5pt}
\resizebox{\linewidth}{!}{
\begin{tabular}{c|c|c|c} 
\hline\hline
System                                                              & On The Fly     & VL      & All     \\ 
\hline\hline
BUT-2022 Wav2vec2.0 + fMLLR + xvectors \cite{baskar2022speaker}                              & \cmark   & 57.72 & 22.83  \\
Nagoya Univ.-2022 WavLM \cite{violeta2022investigating}  & - & 71.50 & 51.80 \\
FAU-2022 Cross-lingual XLSR + Conformer \cite{hernandez22_interspeech} & - & 62.00 & 26.10 \\
JHU-2023 DuTa-VC (Diffusion) + Conformer \cite{wang23qa_interspeech} & - & 63.70 & 27.90 \\
CUHK-2024 HuBERT + sys. comb. \cite{10584335}  & \xmark & 50.70 & 20.56  \\
CUHK-2024 Wav2vec2/HuBERT + GAN Data Aug. + sys. comb. \cite{10447702}  & \xmark & 46.47 & 16.53  \\
CUHK-2024 DA + SVR adapt + sys. comb. \cite{geng2024homogeneous} & \cmark & 57.33 & 23.33  \\
\textbf{HuBERT/WavLM + MoE adapt. + sys. comb. (Sys. 13, Table \ref{tab:main_2}, ours)} & \cmark & \textbf{46.77} & \textbf{16.35} \\
\hline\hline
\end{tabular}
}
\label{table_sota}
\vspace{-0.4cm}
\end{table}

\begin{table}[h]
    \setlength\tabcolsep{2pt}
    \centering
    \caption{Performance (WER\%) of HuBERT using batch-mode MoE-based speaker adaptation under different configurations of \textbf{domain knowledge} (``Know.'') integration and \textbf{KL loss}, as well as on-the-fly MoE-based speaker adaptation with and without \textbf{attentive pooling} (``Atten. Pool.''). Different \textbf{expert types} are investigated.}
    \vspace{-0.2cm}
    \resizebox{\linewidth}{!}{
    \begin{tabular}{c|c|c|c|c|c|c|c|c|c} 
\hline\hline
\multirow{2}{*}{Sys.} & \multirow{2}{*}{\begin{tabular}[c]{@{}c@{}}Expert\\Init.\end{tabular}}                      & \multirow{2}{*}{\begin{tabular}[c]{@{}c@{}}KL\\Loss\end{tabular}} & \multirow{2}{*}{\begin{tabular}[c]{@{}c@{}}Class.\\Task\end{tabular}} & \multirow{2}{*}{\begin{tabular}[c]{@{}c@{}}Domain\\Know.\end{tabular}} & \multirow{2}{*}{\begin{tabular}[c]{@{}c@{}}\# of \\Expert\end{tabular}} & \multirow{2}{*}{\begin{tabular}[c]{@{}c@{}}Atten.\\Pool.\end{tabular}} & \multirow{2}{*}{\begin{tabular}[c]{@{}c@{}}On The\\Fly\end{tabular}} & \multicolumn{2}{c}{WER (\%)}  \\ 
\cline{9-10}
                      &                                                                                             &                                                                   &                                                                       &                                                                            &                                                                         &                                                                      &                                                                        & VL / L / M / H & All               \\ 
\hline\hline
1                     & random                                                                                      & \xmark                                             & \xmark                                                 & -                                                                          & 5                                                                       &   -                                              & \xmark                                                                     & 57.97 / 21.96 / 11.57 / 2.61 & 21.07              \\ 
\hline
2                     & \multirow{9}{*}{\begin{tabular}[c]{@{}c@{}}Init.\\from\\adapt.\\training\\RAB\end{tabular}} & \xmark                                             & \xmark                                                 & \multirow{4}{*}{Severity}                                                  & \multirow{4}{*}{5}                                                      & \multirow{2}{*}{-}                               & \multirow{2}{*}{\xmark}                                                      & 57.01 / 21.33 / 10.98 / 2.73 & \textbf{20.63}              \\ 
\cline{1-1}\cline{3-4}\cline{9-10}
3                     &                                                                                             & \multirow{8}{*}{\cmark}                            & \xmark                                                 &                                                                            &                                                                         &                                                                      &                                                                        & 55.72 / 20.74 / 10.16 / 2.82 & \textbf{20.08}             \\ 
\cline{1-1}\cline{4-4}\cline{7-10}
4                     &                                                                                             &                                                                   & \multirow{7}{*}{\cmark}                                &                                                                            &                                                                         & -                                                & \xmark                                                                     & 55.61 / 20.54 / 10.29 / 2.58 & \textbf{19.94}               \\
5                     &                                                                                             &                                                                   &                                                                       &                                                                            &                                                                         & \cmark                                                & \cmark                                                  & 55.61 / 20.92 / 10.35 / 2.54 & 20.04           \\ 
\cline{1-1}\cline{5-10}
6                     &                                                                                             &                                                                   &                                                                       & \multirow{3}{*}{\begin{tabular}[c]{@{}c@{}}Severity,\\Gender\end{tabular}} & \multirow{3}{*}{10}                                                     & -                                                & \xmark                                                                       & 54.76 / 20.93 / 9.88 / 2.48  & \textbf{19.75}             \\
7                     &                                                                                             &                                                                   &                                                                       &                                                                            &                                                                         & \cmark                                                & \cmark                                                   & 54.42 / 21.02 / 9.88 / 2.59  & \textbf{19.74}               \\
8                     &                                                                                             &                                                                   &                                                                       &                                                                            &                                                                         & \xmark                                                & \cmark                                                 &   55.13 / 21.06 / 10.67 / 2.58                                                &    20.05              \\ 
\cline{1-1}\cline{5-10}
9                     &                                                                                             &                                                                   &                                                                       & \multirow{2}{*}{Speaker}                                                   & \multirow{2}{*}{29}                                                     & -                                                & \xmark                                                                      & 55.76 / 20.35 / 9.59 / 2.72  & 19.84              \\
10                    &                                                                                             &                                                                   &                                                                       &                                                                            &                                                                         & \cmark                                                & \cmark                                                   & 54.12 / 21.03 / 10.06 / 2.71 & 19.75              \\
\hline\hline
\end{tabular}
    }
    \label{tab:ablation}
    \vspace{-0.3cm}
\end{table}


\vspace{-0.2cm}

\vspace{-0.2cm}
\subsection{Ablation Studies}\label{sec:ablation}
\vspace{-0.1cm}
Table \ref{tab:ablation} presents the results of ablation studies on both batch-mode and on-the-fly MoE-based speaker adaptation on the HuBERT model. Several trends can be observed:
\par
\noindent
\textbf{For batch-mode:} \textbf{1)} Regarding the utilization of domain knowledge, initializing each expert with severity-specific adapter parameters from adaptive training (Sys. 2 vs. 1) and incorporating a severity classification task (Sys. 4 vs. 3) both lead to performance improvements; and \textbf{2)} the incorporation of KL loss produces large WER reductions (Sys. 3 vs. 2).
\par
\noindent
\textbf{For on-the-fly:} \textbf{1)} the on-the-fly MoE system with severity-gender experts outperforms the system with severity experts, while achieving comparable performance to the system with speaker-level experts (Sys. 7 vs. 5, 10); and \textbf{2)} the attentive pooling module produces better performance compared to the simple temporal average pooling (Sys. 7 vs. 8).

\vspace{-0.1cm}
\begin{table}[h]
\centering
\caption{Performance of on-the-fly MoE-based adaptation of WavLM with or without the speaker-level round-robin setting. $\ast$ denotes statistically significant improvements against System 1.
}
\vspace{-0.2cm}
\resizebox{\linewidth}{!}{
\begin{tabular}{c|c|c|c|c} 
\hline\hline
Sys. & MoE Adapt. & Round-robin & VL / L / M / H & All  \\ 
\hline\hline
1    &   \xmark       &  \xmark         & 56.49~~ / 22.76~ / 11.57~ / 2.89~                     & 21.06~     \\ 
\hline
2    &  \cmark        &  \xmark         & 53.92$^\ast$ / 21.00$^\ast$ / 10.02$^\ast$ / 2.77~ & 19.72$^\ast$     \\
3    &  \cmark        &  \cmark         &   54.63$^\ast$ / 21.80$^\ast$ / 11.51~ / 3.02~             &  20.21$^\ast$    \\
\hline\hline
\end{tabular}
}
\label{table_zeroshot}
\vspace{-0.3cm}
\end{table}

\vspace{-0.2cm}
\subsection{Analysis}
\vspace{-0.1cm}

\par
\noindent
\textbf{Zero-shot Adaptation.} To evaluate zero-shot performance of the on-the-fly MoE-based speaker adaptation, speaker-level round-robin experiments \cite{8283503} are conducted. Specifically, for each test dysarthric speaker $s$, we exclude their data from the training set before speaker adaptation. As shown in Table \ref{table_zeroshot}, the zero-shot on-the-fly MoE-based adaptation significantly outperforms the SI baseline (Sys. 3 vs. 1), even though the SI model is trained on data containing speakers from the test set.

\begin{figure}[h]
    \vspace{-0.2cm}
    \centering
    \includegraphics[width=0.4\textwidth]{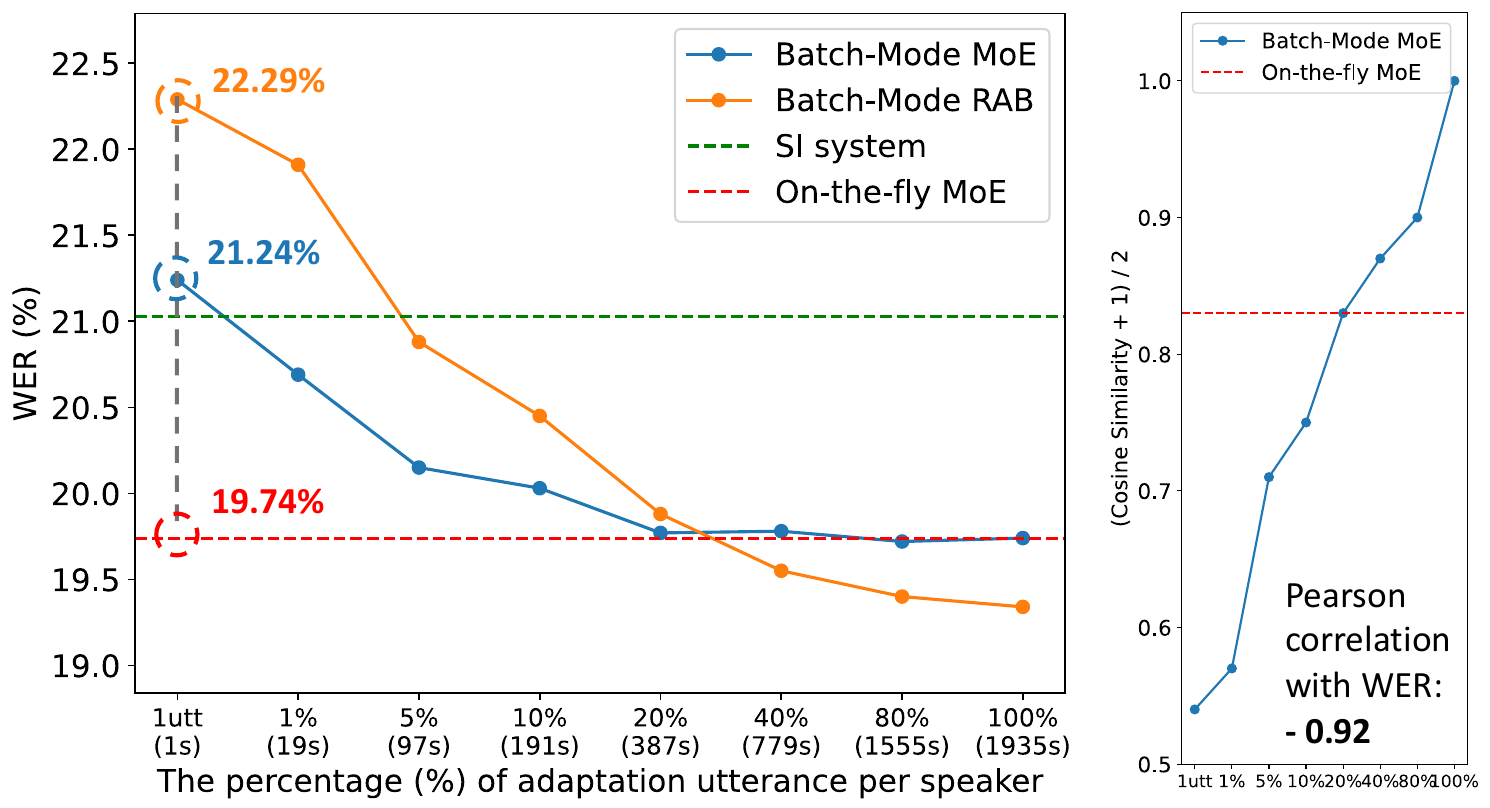}
        \label{fig:hubert_wavlm_per}
    \vspace{-0.2cm}
    \caption{WER and cosine similarity on HuBERT systems with varying amounts of speaker adaptation data.}
    \label{fig:allper}
    \vspace{-0.6cm}
\end{figure}

\begin{figure}[h]
    \centering
    \includegraphics[width=0.45\textwidth]{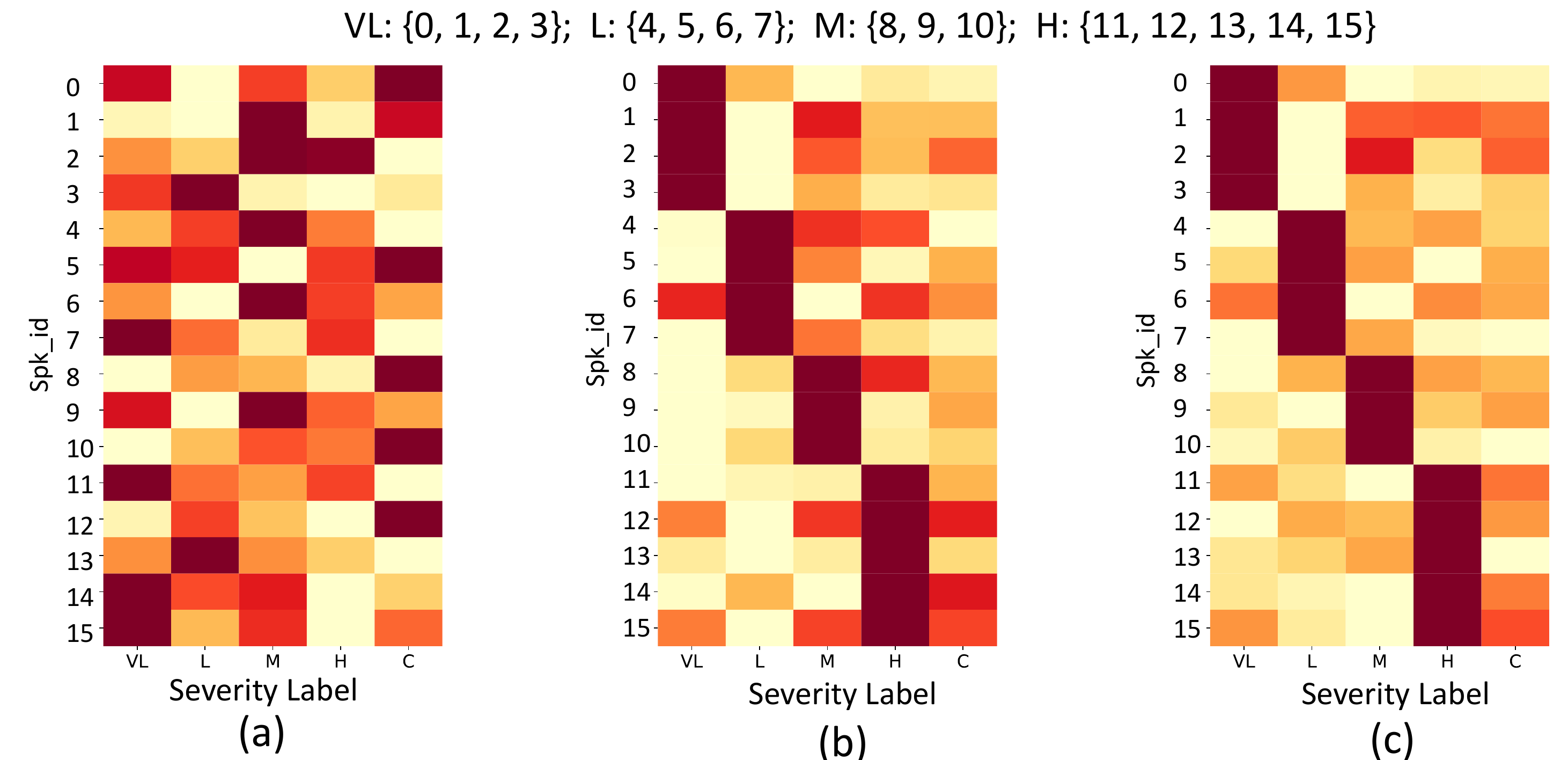}
    \vspace{-0.3cm}
    \caption{Heatmap visualization of routing parameters under varying  settings on HuBERT: \textbf{a)} batch-mode without domain knowledge, \textbf{b)} batch-mode and \textbf{c)} on-the-fly with domain knowledge. ``Severity Label: \{Spk\_ids\}'' is given at the top of Figure.}
    \label{fig:heatmap}
    \vspace{-0.3cm}
\end{figure}

\par
\noindent
\textbf{Low Processing Latency.} The left part of Fig. \ref{fig:allper} illustrates the WER variation of the HuBERT model across different speaker-level data quantities for various adaptation approaches. With just one utterance, the proposed on-the-fly MoE-based adaptation (red line) achieves considerable WER reductions of up to \textbf{2.55\%} absolute (\textbf{11.44\%} relative) compared to both batch-mode RAB and MoE approaches. The right sub-figure illustrates the cosine similarity between routing parameters obtained from partial versus complete data in the batch-mode MoE system. A strong negative correlation (Pearson coefficient = -0.92) between WER and cosine similarity is observed.

\par
\noindent
\textbf{Domain Knowledge Benefits.} Heatmap visualization of routing parameters under different settings on HuBERT in Fig. \ref{fig:heatmap} shows that both on-the-fly and batch-mode MoE-based adaptation incorporating domain knowledge exhibit more consistent and interpretable speech impairment severity-centric features than those without domain knowledge (Fig. \ref{fig:heatmap}(b), (c) vs. (a)).

\vspace{-0.2cm}
\section{Conclusion}
This paper presents a novel on-the-fly MoE-based speaker adaptation for SSL pre-trained SFMs on dysarthric speech. Feature-driven routing networks are designed to produce homogeneous SD routing parameters on the fly, thereby facilitating zero-shot and real-time speaker adaptation. 
Incorporating domain knowledge, such as severity and gender, ensures diverse expert specialization and comprehensive MoE coverage. Experiments on UASpeech shows that the proposed on-the-fly MoE-based adaptation approaches produce up to 1.34\% absolute (6.36\% relative) WER reductions over unadapted SFMs. Comparable WER performance and RTF speedup ratios of 7 times are also obtained over batch-mode adaptation. Heatmap visualization further demonstrates the interpretability of the proposed methods.

\section{Acknowledgements}
This research is supported by Hong Kong RGC GRF grant No. 14200220, 14200021, 14200324, TRS T45-407/19N, Innovation Technology Fund grant No. ITS/218/21, 
the project of China Disabled Persons Federation (CDPF2023KF00002), Basic Research Project of ISCAS (ISCAS-JCMS-202306), Youth Innovation Promotion Association CAS Grant (2023119), and Guangzhou CASTF project (2022MZK02).

\bibliographystyle{IEEEtran}
\bibliography{mybib}

\end{document}